\documentclass[conference]{IEEEtran}

\usepackage{epsfig}
\usepackage{multirow}
 \usepackage{tabu}
\usepackage{microtype} 
\usepackage{booktabs}  
\usepackage{url}
\usepackage{paralist}
\usepackage{subfig}
\usepackage{todonotes}
\usepackage{amsmath}
\usepackage{url}
\usepackage{amssymb}
\usepackage{pifont}
\newcommand{\cmark}{\ding{51}}%
\newcommand{\xmark}{\ding{55}}%
\usepackage{hhline}
\usepackage{makecell}
\usepackage{array}
\usepackage{threeparttable}
\usepackage{boldline}

\usepackage{color}

\begin{document}
	
\date{}

\title{SAT-based Reverse Engineering of Gate-Level Schematics using Fault Injection and Probing}

\author{\IEEEauthorblockN{Shahrzad Keshavarz\IEEEauthorrefmark{1}, Falk Schellenberg\IEEEauthorrefmark{2}, Bastian Richter\IEEEauthorrefmark{2}, Christof Paar\IEEEauthorrefmark{2}\textsuperscript{,}\IEEEauthorrefmark{1}
		and Daniel Holcomb\IEEEauthorrefmark{1}
		\IEEEauthorblockA{\IEEEauthorrefmark{1}University of Massachusetts Amherst, USA
			\\ \{skeshavarz, dholcomb\}@umass.edu 
		}
		\IEEEauthorblockA{\IEEEauthorrefmark{2}Horst G{\"o}rtz Institute for IT Security,  Ruhr-Universit{\"a}t Bochum, Bochum, Germany\\{\{firstname.lastname\}@rub.de}
	}}
\\[-3.84ex]
}
\maketitle

\begin{abstract}
Gate camouflaging is a known security enhancement technique that tries to thwart reverse engineering by hiding the functions of gates or the connections between them. A number of works on SAT-based attacks have shown that it is often possible to reverse engineer a circuit function by combining a camouflaged circuit model and the ability to have oracle access to the obfuscated combinational circuit. Especially in small circuits it is easy to reverse engineer the circuit function in this way, but SAT-based reverse engineering techniques provide no guarantees of recovering a circuit that is gate-by-gate equivalent to the original design. In this work we show that an attacker who doesn't know gate functions or connections of an aggressively camouflaged circuit cannot learn the correct gate-level schematic even if able to control inputs and probe all combinational nodes of the circuit. We then present a stronger attack that extends SAT-based reverse engineering with fault analysis to allow an attacker to recover the correct gate-level schematic. We analyze our reverse engineering approach on an S-Box circuit.

\end{abstract}

\section{Introduction}
\label{sec:introduction}
Gate camouflaging is a technique that has attracted the attention of chip designers in recent years. Camouflaging seeks to hide the true structures of the chip so that imaging-based reverse engineering cannot easily recover the details of the implemented design. The purposes of camouflaging include IP protection and preventing targeted attacks. The related work section of this document describes some of the different camouflaging mechanisms that exist in academia and industry.

A number of attacks exist against camouflaging including the SAT attack which is based on Boolean satisfiability solving. In this attack, a reverse engineer uses an uncertain model of the design, together with a functional instance of the chip as an oracle, to discover a set of tests that will reveal the exact logic function of the design. The SAT attack extracts the correct function of the design, but is unable to make any claim regarding whether it has recovered a design that matches the gate-level functions of the obfuscated one, or another gate-level schematic that is overall functionally equivalent to the obfuscated design. In this work, we present a stronger SAT attack for small circuits that makes the following contributions:


\begin{itemize}
        \item We show how an attacker with probing and fault injection capability can use SAT-based reverse engineering to guide his decisions about which faults to apply and which nodes to probe.
	\item We propose a new SAT-based reverse engineering formulation that can solve for unknown connections while restricting the search to acyclic networks and avoiding combinational loops that can thwart SAT attacks.
	\item We show that fault injection and probing provide additional discriminating factors in reverse engineering that can help SAT attacks to recover schematics that are equivalent to the target on a gate-by-gate basis, instead of merely functionally equivalent in traditional SAT attacks.
\end{itemize}

\section{Related Work and Background}


Imaging-based invasive reverse engineering works by decapsulating the chip, imaging and removing each layer in succession, and then using the images to reconstruct the circuit schematic.

A countermeasure against imaging-based reverse engineering is the use of various camouflaged gates or camouflaged interconnects. Camouflaged components are ones in which different functions are implemented by features that are indistinguishable to the reverse engineer, so that function cannot be inferred from appearance. Camouflaged gate libraries use hard-to-observe structural techniques to differentiate the gate functions~\cite{syphermedia-library,cocchi-14}, use functionality that can be controlled without structural differences via transistor doping~\cite{becker-13,shiozaki-14,malik-obfusgate,iyengar-15,collantes-16}, use conducting and non-conducting interconnects, or use secret key inputs that control the design functionality with a structure \cite{KeshavarzSRAM} that cannot be distinguished by the attacker once the chip is delayered~\cite{chen-2015-dummyWire}. Though camouflaging is a promising hardware security enhancement technique, it comes with additional overheads in the chip area, power consumption, and fabrication cost, and there is always a trade-off between the security and overheads~\cite{chakraborty-09,rajendran-13}. 

\subsection{SAT Attacks}
SAT attacks are based on principle of finding discriminating input vectors, which are input vectors that can eliminate at least one additional circuit function hypothesis once the corresponding output vector is known. Once no further discriminating vectors can be found, it means that no further circuit functions can be ruled out by any tests, and therefore the current set of discriminating inputs is sufficient to uniquely identify the circuit function. Techniques from oracle-guided synthesis~\cite{jha2010oracle} are used in SAT-based attacks to reverse engineer gate camouflaging or logic encryption~\cite{elmassad-15,subramanyan-15}.

It is important to note that a circuit reverse engineered by oracle-guided synthesis is only guaranteed to be functionally equivalent to the obfuscated circuit, and there is no assurance that it will match the obfuscated circuit on a gate-by-gate basis. Ensuring gate-by-gate equivalence to the obfuscated circuit is generally impossible because the attack only has information about the inputs and outputs. Designs recovered through oracle-guided synthesis may therefore be unsuitable for certain classes of side-channel or fault injection attacks that require knowing the states of all combinational circuit nets. In this paper, we propose a SAT-based de-obfuscation technique that assumes very little knowledge about the obfuscated circuit connections or gates, yet still attempts to reconstruct the exact gate-level schematic of the obfuscated circuit.

\subsection{Attacker Model}
The attacker model we consider in this work represents an adversary that is trying to reverse engineer a circuit from the backside. This scenario may arise in chips with anti-tamper mechanisms that prevent delayering to learn the interconnections of each metal layer. From the backside, the adversary has a very limited knowledge of the circuit as listed below:

 \textbf{Connections:} All connections in the circuit are unknown.
  This means that any gate input in the circuit could be connected to the output of any other gate in the circuit. 
 
 \textbf{Gate inputs/outputs:} Each gate has a single output, and the output pin of the gate can be identified, yet the adversary cannot see what the gate output connects to. The adversary can know how many inputs each gate has, but cannot know which signals (primary inputs or outputs of other gates) are driving them. If the number of inputs to each gate cannot be determined, the attacker can be conservative and overestimate the number of inputs to each gate. 


\textbf{Gate functions:} Our model considers that the attack may know nothing about the gate functions. That is, a gate with $n$ inputs can implement any of  $2^{2^n}$ possible functions.


\vspace{5pt}
The assumed attacker capabilities in this work are as described below:
  
 \textbf{Circuit inputs/outputs:} Attacker has a working circuit instance, and can apply the desired inputs to the circuit and observe the outputs.
 The circuit instance used to correctly map input vectors to output vectors is called the ''oracle''.
In case the circuit is part of an encryption hardware, the
attacker can control the input to the encryption hardware and knows (or is able to
set) an internal secret key. This enables calculating any intermediate
values that might occur during computation (the primary outputs of our target circuit).


\textbf{Probes: } At some points in the work, the attacker is allowed to probe the value of arbitrary gate outputs. In this setting, the attacker still has no knowledge of connectivity and hence doesn't know what else is being driven by the node that is probed. 
Due the nature of probing, it is not possible to probe the value of gate inputs.

\textbf{Fault Injection: } At some points in the work, the attacker is allowed to inject faults using a laser.

\section{SAT Formulation for Unknown Gates and Connections}
\label{sec:sat-formulation}
\begin{figure}[!ht]
	\centering
		\includegraphics[width=0.78\columnwidth]{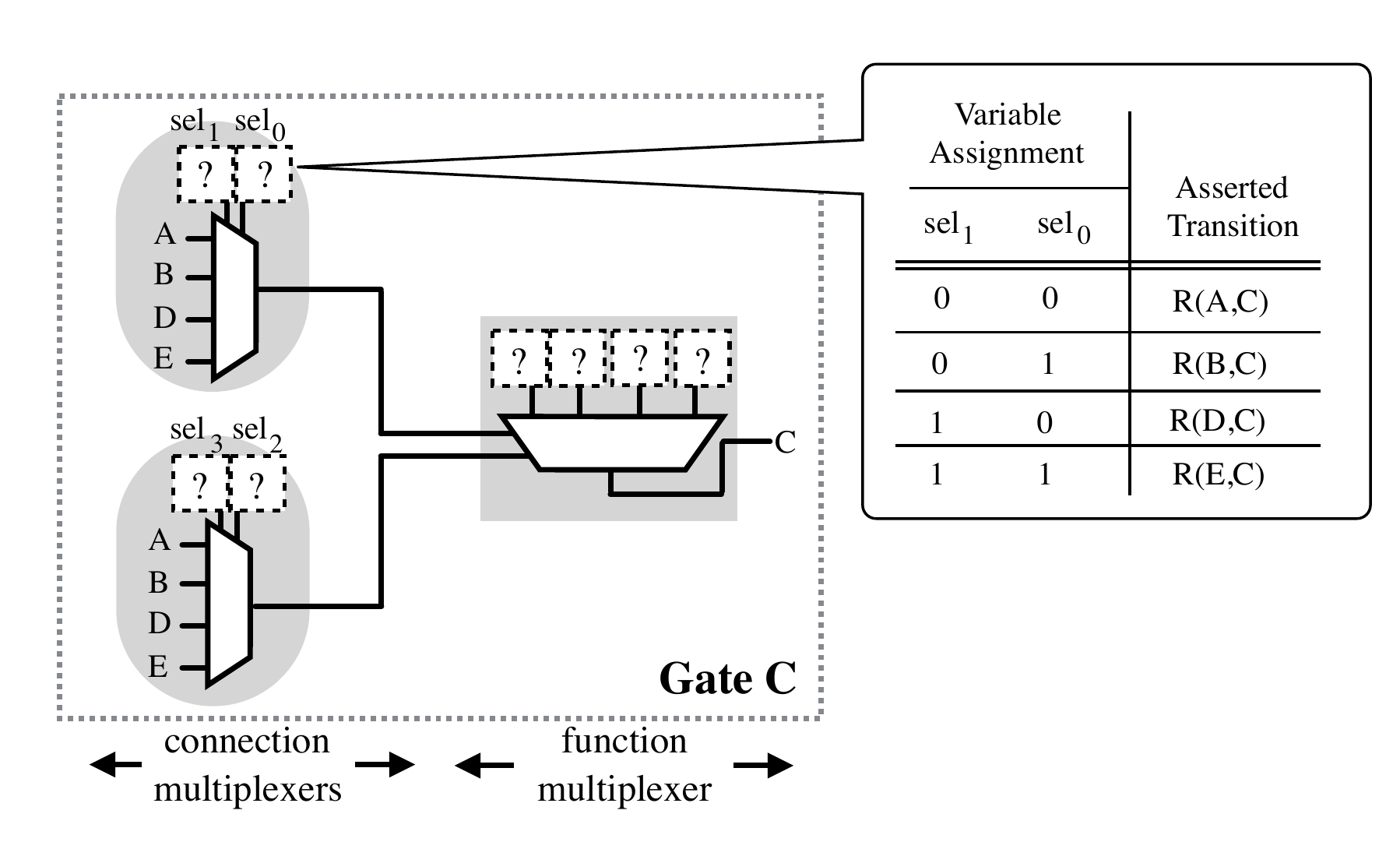} 
		\label{camcircuit}
	\caption{An example of the proposed gate model. Depending on the values of the configuration variables, this model allows each gate input to be driven from any node, and allows the gate to implement any possible logic function over its inputs.}
	\vspace{-10pt}
	\label{fig:ckt_and_model}
\end{figure}

We demonstrate the modeling of connections and gate functions using the example shown in Fig.~\ref{fig:ckt_and_model}. In this example, an unknown 2-input gate has output node $C$ and thus is denoted as gate $C$. The gate exists within a circuit having 5 nodes ($A$, $B$, $C$, $D$ and $E$). The uncertainty about logic function of gates and uncertainty about wiring connections are both translated into Boolean configuration variables (shown as white dotted-line boxes) that are connected to multiplexers. The values of configuration variables are unknown, and the SAT solver's task is to find them. 


Since nothing about the connection of gates are known to the attacker, multiplexers are added that are responsible to select which node in the circuit is connected to each input of the gate. For example, in Fig. \ref{fig:ckt_and_model}, since the gate has output $C$, the connection multiplexers choose from the other four nodes of the circuit ($A$, $B$, $D$ and $E$) to determine which is connected to each of the gate's inputs. In a circuit with $N$ nodes, the connection multiplexers are therefore $(N-1)$-to-1 input multiplexers, as they can select any other node in the circuit except for that gate's own output (node $C$). In some cases, as will be shown later, certain connections can be ruled out and the number of multiplexer inputs would reduce accordingly. 

To keep track of the connectivity between gates, as will be needed later to ensure that the solver only considers acyclic networks, we define transition relation predicates for all pairs of gates. If there is a connection from output of gate $A$ (node $A$) to one of the inputs of gate $C$ (that has output node $C$), the predicate $R(A,C)$ will be 1 and otherwise it will be 0.
In Fig. \ref{fig:ckt_and_model}, predicate $R(A,C)$ is true if and only if the configuration variables for the connection multiplexer connect the output of gate $A$ to an input of gate $C$; therefore, $R(A,C)$ is true whenever $sel_1sel_0=00$ or $sel_3sel_2=00$.

The second type of multiplexer employed is for choosing the function of the gate based on the selected inputs from the connection multiplexers. The function multiplexer can be regarded as implementing the truth table of the gate function, choosing which combination of input values should result in which binary value on the gate's output. For a gate of $n$ inputs, the function multiplexer would be a $2^n$ to 1 multiplexer.
Note that our model puts no restrictions on the function of the gates.
However, if the attacker has knowledge of the gate library used, he can put restrictions on the configuration variables that determine the gate's function. For example, in figure \ref{fig:ckt_and_model}, if the attacker knows that the 2-input gate could only be $NAND$ or $NOR$, then he can restrict the multiplexer's input values to ''1110'' (for $NAND$ gate) and ''1000'' (for $NOR$ gate) by adding clauses to the SAT problem to disallow all other combinations. 
\vspace{-5pt}

\section{Learning from Voltage Probing}
\label{sec:probing}
\label{sec:feasibleInputs}

Adding more constraints and known facts to the SAT problem can make it easier to solve. One approach that can help the attacker with reverse engineering is a semi-invasive technique called laser voltage probing (LVP) \cite{Lohrke2016, wang2017probing}. In laser voltage probing, the target transistors are illuminated and the signal values are inferred based on the measured emitted light. Two broad classes of voltage probing are frontside and backside. The frontside of the chip is the side of metal layers while the backside is the side of the substrate. With the growing number of metal layers on the frontside, backside probing may seem more promising as it keeps the metal layers intact and preserves the proper functionality of the circuit \cite{kindereit2014fundamentals}.
 Preparing the chip for frontside probing requires decapping the chip by removing epoxy and
 blocking metal layers to access the internal signals or transistors while backside probing only requires simple thinning and polishing from the back \cite{van2011practical, boit2008physical}. 
The ability to probe node values enhances the attacker's observability.


Having access to the value of internal signals can also help the SAT solver to make inference about the possible connections between gates. Even when gate functions are unknown, it is known due to the nature of circuits that each gate instance must implement a deterministic Boolean function; in other words, any gate must always map the same gate input value to the same gate output value.

\begin{table}%
  \centering
  \subfloat[][Probed node values.\label{tab:probed}]{
  	\scalebox{0.8}{
	\begin{tabular}{l|l|l|l|l}
\multicolumn{1}{l|}{input} & 		\multicolumn{4}{c}{circuit nodes}  \\ 	
vector & A & B & C & X \\ \hline \hline
0000      & 1 & 1 & 0 & 0 \\ \hline
0001      & 0 & 1 & 0 & 1 \\ \hline
0010      & 1 & 1 & 1 & 0 \\ \hline
0011      & 0 & 0 & 1 & 0 \\ \hline
0100      & 0 & 1 & 0 & 1 \\
	\end{tabular}
}
}%
  \quad
  \subfloat[][Gate truth table for different connections.\label{tab:tt}]{
  	  	\scalebox{0.82}{
	\begin{tabular}{l|l c l|l c l|l}
AB & X   && AC & X   && BC & X     \\ \cline{1-2} \cline{4-5} \cline{7-8} \cline{1-2} \cline{4-5} \cline{7-8}
00 & 0   && 00 & 1,1 && 00 &       \\ 	\cline{1-2} \cline{4-5} \cline{7-8}
01 & 1,1 && 01 & 0   && 01 & 0     \\ \cline{1-2} \cline{4-5} \cline{7-8}
10 &     && 10 & 0   && 10 & 0,1,1 \\ \cline{1-2} \cline{4-5} \cline{7-8}
11 & 0,0 && 11 & 0   && 11 & 0     \\ 
	\end{tabular}
}
}
  \caption{Example showing that probed values can rule out certain connections. One of the three possible node pairings is non-deterministic and will be rulled out by SAT solver.}
    \vspace{-15pt}
  \label{tab:probing_example}
\end{table}

As a demonstration of how probing can eliminate some candidate connections, consider the example of Tab.~\ref{tab:probing_example} that shows the values of selected nodes when five different primary input values are applied to a circuit. Assume in this case that the attacker knows that node $X$ is the output of a 2-input gate and nodes $A$, $B$, and $C$ are other nodes in the circuit. Without knowing the connections of the circuit, the attacker knows only that the inputs to the 2-input gate that produces $X$ are either ($A,B$), ($A,C$), or ($B,C$).
For each primary input that is applied to the circuit, probing the values of node X and each of the possible gate input connection pairs make up different cases as shown in Tab.~\ref{tab:tt}.
Looking at these truth tables, we can see that it is impossible for the gate input connections to be ($B,C$), because $X$ takes different values in the three vectors that induced ($B,C$) to have the values (1,0). Input combinations ($A,B$) and ($B,C$) both imply a consistent (deterministic) function for the gate, so neither of these can be ruled out. Note that our pair notation is not ordered; in other words, ($n_i, n_j$) = ($n_j, n_i$).



Using probed values to rule out infeasible input combinations leads to, for each gate, a set of feasible input pairs. If the set of nodes in the circuit is denoted as $N$, for each node $n_x \in N$ that is the output of a 2-input gate a set of feasible input pairings ($F(n_x)$) can be calculated as shown below, where $n_i^j$ is the value of node $n_i \in N$ when the $j^{th}$ input vector is applied to the circuit.



\vspace{-10pt}

 \begin{equation}
 \label{eq:infeasibleset}
\begin{small}
\begin{aligned}
F({n_x}) \hspace{-3pt} := \hspace{-2pt}
(n_y, n_w) \hspace{-2pt} \in \hspace{-2pt} N^2  \hspace{-2pt} \mid  \hspace{-2pt}	\left( (n_y^i,n_w^i) \hspace{-1pt}  = \hspace{-1pt}  (n_y^j,n_w^j) \right) \hspace{-2pt} \Rightarrow \hspace{-3pt} \left( n_x^i \hspace{-2pt} = \hspace{-2pt} n_x^j \right)
\end{aligned}
\end{small}
\end{equation}

\section{Learning from Fault Injection}
\label{sec:fault-injection}

Our formulation incorporates the use of fault injection in the reverse engineering process.
The use of fault injection is important when trying to reverse engineer gate-level schematics because input/output observations and probing can be insufficient to uniquely recover the implementation. In our setting, the attacker targets specific nodes as instructed by the SAT solver, but performs each fault injection on a node without knowing the function of any gates or their connections. 


In laser fault injection, the attacker can use a setup
that is very similar to that used for probing~\cite{schellenberg-16}. However, instead of measuring the reflected light, he chooses
wavelength and energy of the laser pulse so that the photoelectric
effect occurs. When focusing the laser beam at a transistor node, an electric current is generated. The induced current might charge
or discharge the output of the gate, depending on whether the targeted transistor is a PMOS in the gate's 
pull-up network or an NMOS in the gate's pull-down network. The ability to inject such single bit
errors has been experimentally verified down to 45nm feature
size \cite{cardis15_lfi}. Given that the duration of laser faults can exceed the clock period, they can be modeled as stuck-at faults in the circuit model.

Masking is an important consideration in fault injection. When the attacker tries to force a 0 or 1 value onto a node for some applied input vector, the induced value will have no effect if it matches the fault-free value of the same node. Similarly, even if the induced value does change the value of the targeted node, it is possible that the changed value may not propagate to the outputs.
Cases where an induced fault changes the output are perhaps the most informative in reverse engineering. In these cases, the attacker learns that the fault-free value of the node is opposite the induced value, and learns the specific output value that is caused by the fault. The information learned from different fault injection outcomes is listed in Tab.~\ref{tab:obs}.
We show in the formulation described in section~\ref{sec:sat-lvp} that inference from fault injection can be integrated into a SAT-based reverse engineering framework, and this leave the deductions shown in Tab.~\ref{tab:obs} to be made by the SAT solver.

\begin{table}[]
	\centering
	\caption{
When a fault is applied, a corresponding output observation is made that is either $o_{correct}$ (if it matches the non-faulted circuit output) or $o_{faulty}$ (if it differs from the non-faulted output). The table summarizes the information that is revealed by each outcome.}
	\label{tab:obs}
	  	\scalebox{0.9}{
	\begin{tabular}{l|l|l}
condition & output & information learned \\ \hline
SA-1 & $o_{faulty}$ &  \begin{tabular}[c]{@{}l@{}}fault-free value of target node is 0 AND\\ fault must propagate to observed outputs\end{tabular} \\\hline
SA-0 & $o_{faulty}$ &  \begin{tabular}[c]{@{}l@{}}fault-free value of target node is 1 AND\\ fault must propagate to observed outputs\end{tabular} \\\hline
SA-1 & $o_{correct}$ & \begin{tabular}[c]{@{}l@{}}fault-free value of target node is 1 OR\\ fault does not propagate to outputs\end{tabular} \\\hline
SA-0 & $o_{correct}$ & \begin{tabular}[c]{@{}l@{}}fault-free value of target node is 0 OR\\ fault does not propagate to outputs
	
\end{tabular} \\\hline
	\end{tabular}
}
\end{table}


\section{Extended SAT Formulation}

\label{sec:extended-sat-formulation}

We have previously shown in section \ref{sec:sat-formulation} how to model each gate based on the attacker's knowledge about the circuit. In this section, we first show how to enforce gate levelization in the model to restrict the solver to loop-free circuits, and then we show how to incorporate the additional information from voltage probing and fault injection into the SAT problem so that it can be used by an attacker having these capabilities.
\subsection{Restriction to Acyclic Topologies}
\label{sec:levelize}
The basic SAT formulation given in Sec.~\ref{sec:sat-formulation} allows for cycles in the wiring connections that would not occur in combinational circuits. The possibility of cycles is problematic because cycles allow circuit nodes to become undefined state variables (that are not determined by the circuit inputs). This allows the solver to repeatedly find erroneous discriminating inputs, which in reality are not useful in the reverse engineering process. 


To avoid this problem, we modify the SAT formulation to disallow cycles while still allowing arbitrary acyclic topologies. Our solution for disallowing cycles is to enforce that the SAT solver only find solutions in which the topology can be levelized (i.e. topologically sorted). We solve for the circuit's levelization as part of the same SAT formulation that solves for the gate functions and connections. To do this, we add auxiliary variables (to denote levels) and levelization constraints to the SAT problem. Our proposed levelization-enforcing approach is not only helpful for our problem and assumptions, but also can help making any SAT attack feasible when there is an uncertainty in connections that could otherwise allow a loop in the circuit. For example, it can be used to break the cyclic obfuscation approach in \cite{shamsi2017cyclic} by solving for the actual levels of gates and hence figuring out the dummy connections that would otherwise make the SAT attack impossible.

%

In the conventional levelization definition, levels increase from inputs to outputs.
In our formulation, the levels increase from outputs to inputs, but otherwise the levelization notion is the same. Any gate connected to a primary output should be assigned level 1, and the level of every gate must be exceeded by the levels of the gates providing its inputs. In other words, the level of any gate should be higher than the level of all its fanout gates. 

\begin{figure}[!ht]
	\centering
	\includegraphics[width=0.65\columnwidth]{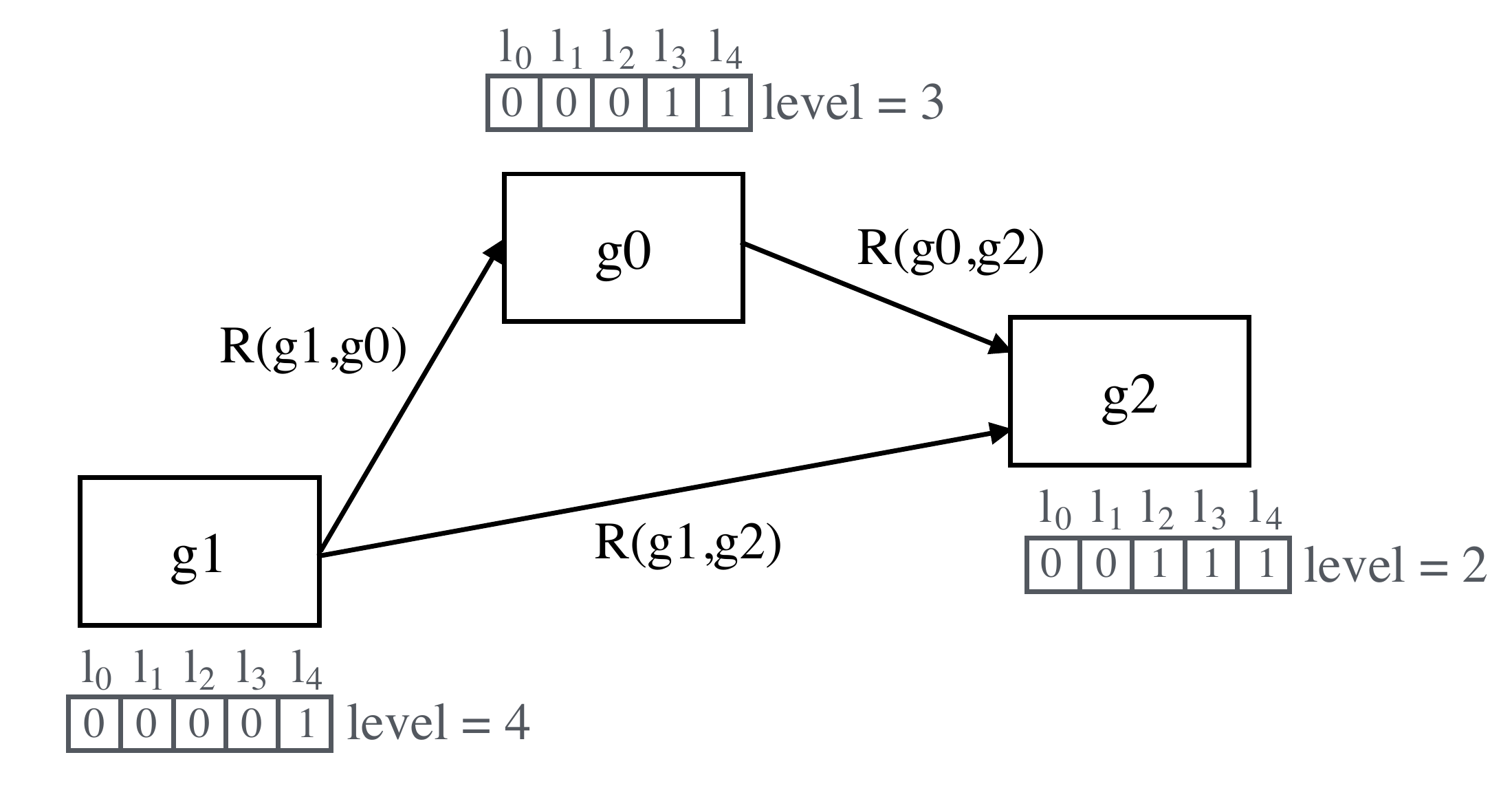}
	\caption{Sample levelization encoding in a circuit}
	\label{fig:loopPrevention}
\end{figure}

{\bfseries Encoding Constraints:}
For each gate $g_j$ in a circuit with $n$ levels, we define a bit-vector of auxiliary variables $(l_0(g_j), l_1(g_j), \dots, l_{n}(g_j))$ to encode the level of the gate. The level of the gate is encoded in a thermometer code style, with a number of 0 values followed by number of 1 values. If bit $l_i(g_j)$ is 0, then gate $g_j$ exceeds level $i$. If bit $l_i(g_j)$ is 1, then the level of gate $g_j$ is less than or equal to $i$. Therefore, the level of the gate can be said to be the left-most bit position in which the value is 1. For example in Fig.~\ref{fig:loopPrevention}, the level of $g_2$ is 2 because $l_2(g_2)$ is the left-most bit position with value of 1. In any legal thermometer coded value, every 0 bit in the vector other than the first must be preceded by another 0 bit, and this is enforced by the encoding invariant shown in eq. \ref{eq:encodingconst}. The first and last bit of the level encoding vector must be 0 and 1 respectively for all gates. 

\vspace{-10pt}
\begin{equation}
 \label{eq:encodingconst}
 \begin{small}
\begin{aligned}
\forall{{\scriptstyle i>0}, g_j}:  
\overbrace{\neg l_i(g_j)}^{level > i}    \Rightarrow \overbrace{\neg l_{i-1}(g_j)}^{level > i-1},\qquad\forall{g_j}: \neg l_0(g_j) \wedge l_n(g_j)
\end{aligned}
\end{small}
\end{equation}

{\bfseries Ordering of Levels:}
For the circuit to be levelized, each gate $g_j$ at level $i$ or greater must get its inputs from gates at level $i+1$ or greater. Using transition predicate $R(a,b)$ (see Sec.~\ref{sec:sat-formulation}) to denote a connection from output of gate $a$ to an input of gate $b$, any legal level-ordering between nodes $a$ and $b$ must obey the ordering constraint in eq.~\ref{eq:orderinconst}. For example in Fig.~\ref{fig:loopPrevention}, $l_1(g_2)=0$ (meaning that gate $g_2$ is level 2 or higher) and we have $R(g_0, g_2)=1$ and $R(g_1,g_2)=1$, indicating that both $g_0$ and $g_1$ fan out to $g_2$. Therefore, the ordering constraint enforces that $l_2(g_0)$ and $l_2(g_1)$ must both be 0, meaning that gates $g_0$ and $g_1$ are at level 3 or higher.

\begin{equation}
\label{eq:orderinconst}
\begin{small}
\begin{aligned}
 \forall{{\scriptstyle i>0},g_j,g_k}: 	\!\!
\left( \overbrace{\neg l_{i-1}(g_j)}^{level \geq i}  \wedge R(g_k, g_j)\right)  
\Rightarrow \overbrace{\neg l_{i}(g_k)}^{fanin\ level \geq i+1} 
\end{aligned}
\end{small}
\end{equation}

\vspace{-8pt}

\subsection{Adding fault injection results to SAT problem}

We now extend SAT attacks to account for the attacker's ability to inject faults.
To incorporate fault injection results into the SAT problem, the attacker can add a new multiplexer to select between faulty or normal values for all nodes (except primary inputs) in his model, and then allow the SAT solver to guide his fault injection trials as will be shown. This new structure is shown with thick lines in Fig. \ref{fig:faultModel}. For a node $n_x$ in the circuit, $injectFault_x$ is a primary input to the model that selects whether or not a fault should be injected on it. Primary input signal $FaultVal$ determines the value that is forced onto the selected node. Node $n_x\_fe$ is the fault-enabled version of the node, which is either the value computed by the circuit for $n_x$, or the value forced onto the node by fault injection. 

 Discriminating inputs produced by the solver now provide the reverse engineer with an input vector to apply to the circuit, as well as a node to fault, and a faulty value to inject on that node. The attacker applies these conditions to the oracle circuit, finds the resulting output vector, and feeds the conditions back into the SAT solver as constraints. The ability to have additional discriminating information through fault injection can allow an attacker to better distinguish between circuit implementations that are overall functionally equivalent. 

\subsection{Adding Voltage Probing to SAT problem}
\label{sec:sat-lvp}

We now extend SAT attacks to also incorporate voltage probing. Voltage probing has the effect of selectively making internal nodes of the circuit visible to the attacker, which in the attacker's model is equivalent to selectively making internal nodes appear as part of the observable output vector. We create an input signal in the model that selectively activates the observability of a node so that, when the SAT solver finds a discriminating input, that input now indicates to the attacker whether any internal signals should be probed when the vector is applied to the oracle. The corresponding output vector and probed value are fed back into the SAT attack as the additional constraints on the circuit configuration variables.

\begin{figure*}[tbh!]
	\centering
	\begin{minipage}[c]{0.22\textwidth}
		\centering
		\subfloat{
			\includegraphics[width=1\textwidth]{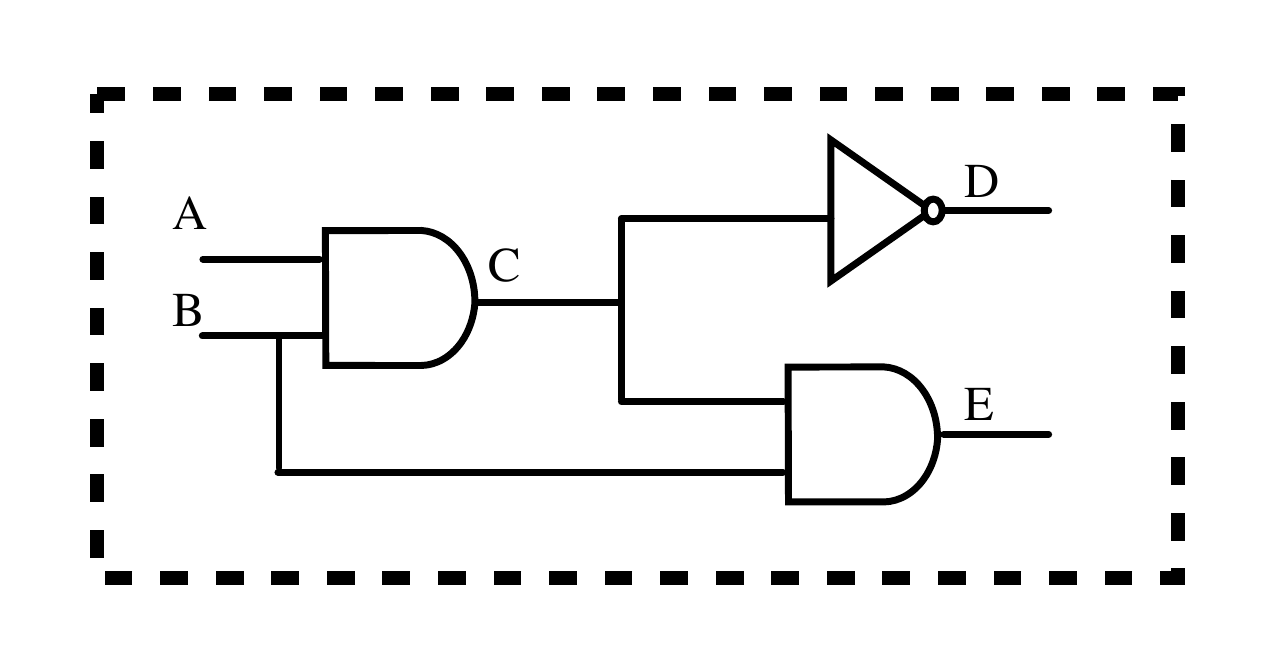}
		}
	\end{minipage}
	\begin{minipage}[c]{0.75\textwidth}
		\centering
		\subfloat{
			\includegraphics[width=1\textwidth]{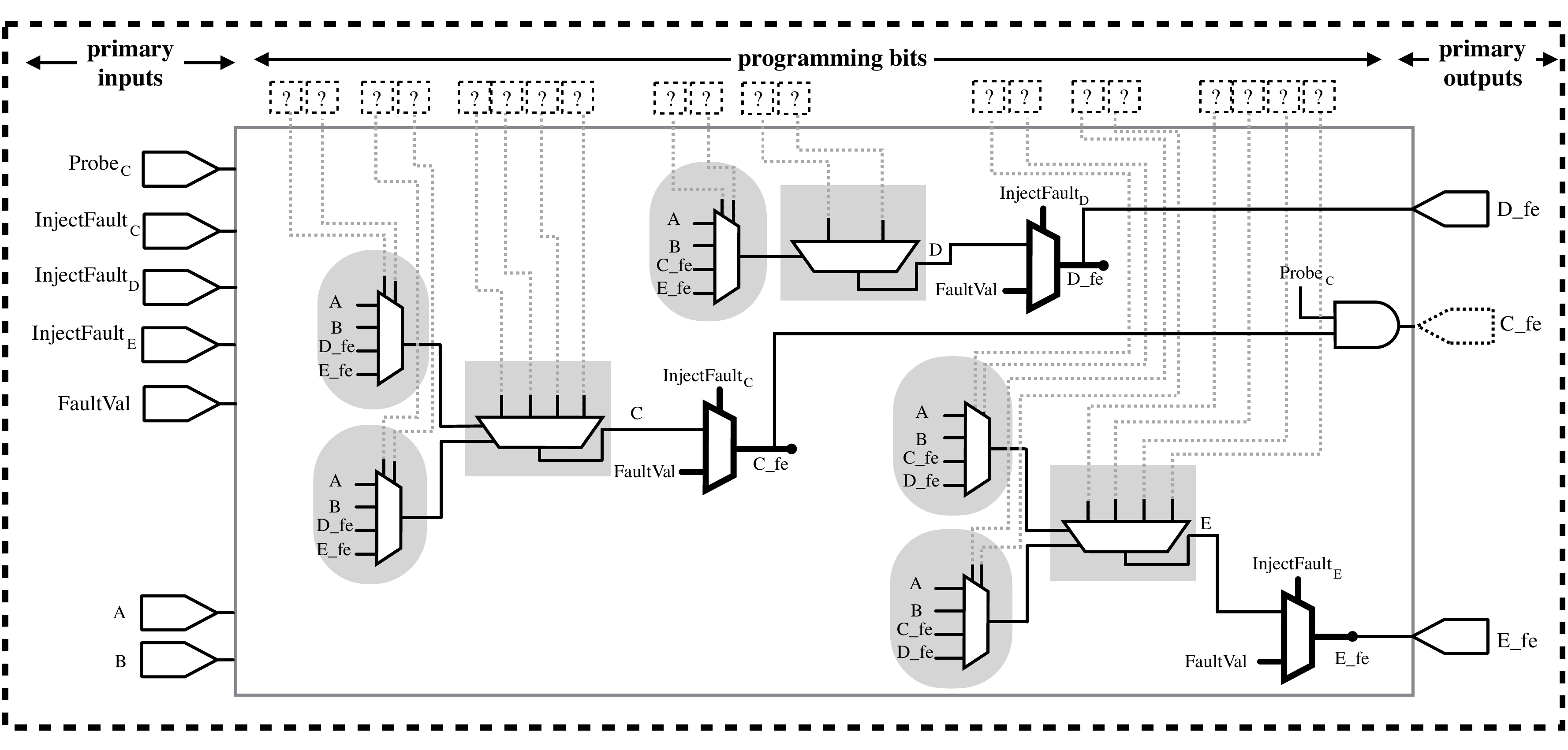}
		}
\end{minipage}

	\caption{The circuit shown on the left would be modeled as the one shown on the right. The multiplexers in thick lines are used to incorporate fault injection and the AND gate controlled with $Probe_C$ signal is used to add voltage probing into the SAT problem. }
	\label{fig:faultModel}
\end{figure*}

Probing is illustrated in Fig.~\ref{fig:faultModel} using a model of a circuit with a single internal node $C$. The newly added input signal $Probe_C$ selects whether the value of node $C$ can be considered when finding a discriminating input vector. Whenever the solver decides to assert the input signal $Probe_C$, then a discriminating vector is one that can produce different values on the primary outputs or on the now-observable signal $C$. Whenever the solver does not assert $Probe_C$, then the approach reverts to the standard SAT attack that discriminates between possible configurations based on the primary outputs only.

\section{Results}
\label{sec:results}
We have evaluated our approach for two small circuits. The first is ISCAS'85 benchmark circuit c17 comprising 6 gates, 4 internal wires, 5 primary inputs and 2 primary outputs; the second is an S-Box from the PRESENT block cipher \cite{bogdanov2007present} comprising 20 gates, 16 internal wires, 4 primary inputs and 4 primary outputs. For our oracle, we simulate the circuit and perform fault simulation using ModelSim. The attack is performed on a fully-camouflaged netlist where all gates and connections are unknown and modeled as explained in Section~\ref{sec:extended-sat-formulation}. We perform the SAT attack using a modified version of a program from existing work~\cite{yu-17}. The results for runtime and number of iterations (i.e. number of discriminating inputs found) before the algorithm terminates are shown in Tab. \ref{tab:res}, and explained in the following subsections. The algorithm terminates when no more discriminating inputs can be found, and at that time, the recovered solution is compared to the original netlist to see whether they are equivalent on a gate-by-gate basis.

\begin{figure}
	\centering

		\includegraphics[width=0.5\columnwidth]{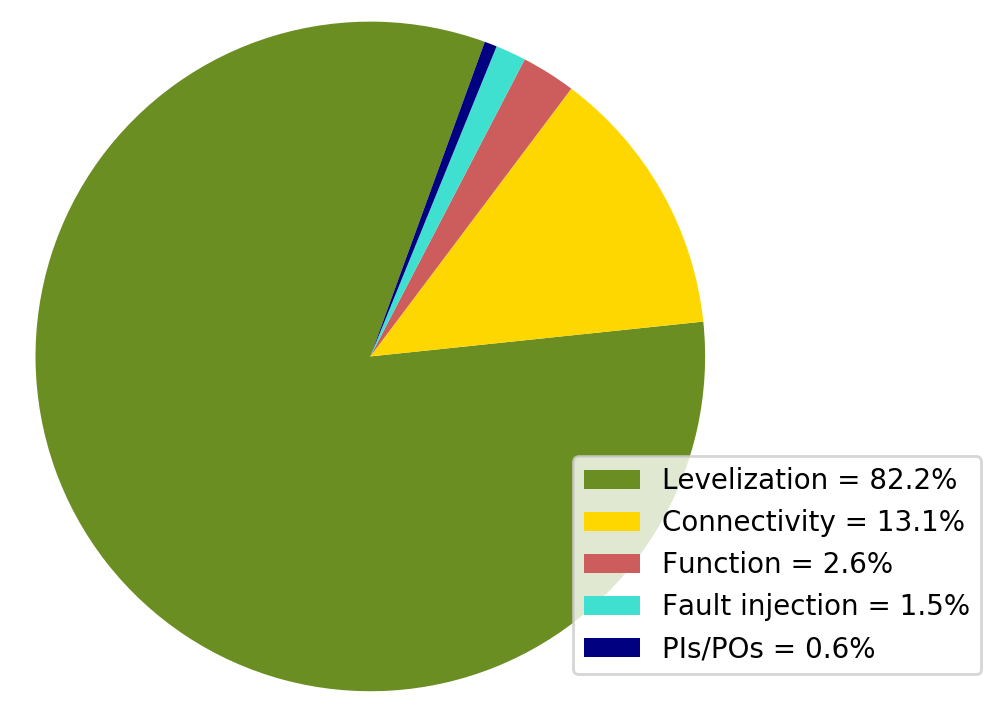}
	\caption{Proportion of CNF variables being used toward each component of model in ISCAS'85 c17 circuit.}
	\label{fig:varRatio}
\end{figure}

\subsection{Distribution of SAT variables}
In all cases, the levelization constraints are found to be necessary for the SAT attack algorithm to terminate successfully (See sec.~\ref{sec:levelize} for details). The majority of the variables and clauses in the SAT problem are used to implement the constraints that enforce levelization.
For circuit c17, Fig.~\ref{fig:varRatio} shows the proportion of SAT variables that are used in each of the following aspects of the formulation: The function variables that help with solving gate functions, the connection variables that are created to solve the gate connections, the levelization variables that are created to enforce levelization in the circuit, the fault injection variables that are used to add fault injection capabilities to the model, and the variables related to circuit's primary inputs and outputs and constraints thereof.

\subsection{Effectiveness of fault injection and probing}
As can be seen in Tab. \ref{tab:res}, the problem is not solved within hours if probing and fault injection are not used. When probing is enabled, but fault injection is not, the algorithm converges to a solution in a timely manner. However, the solution is not unique, and in the case of the S-Box the recovered circuit does not match the structure of the target circuit that is being reverse engineered.
Therefore, probing alone doesn't fulfill our objective of finding solutions that are structurally equivalent to the original netlist on a gate-by-gate basis. 

Using fault injection along with voltage probing in reverse engineering makes it possible to find a unique solution for both circuits. In case of the S-box circuit with fault injection and probing, approximately 800K variables and 5M clauses were generated to solve the problem. This solution is identical to the target circuit in all connections and all gate functions. Note that the formulation that includes both probing and fault injection requires more iterations (more discriminating inputs). This occurs because of the large space of fault injection tests and the need to rule out every possible circuit configuration that is not exactly the same as the target, instead of merely ruling out the configurations that are not functionally identical to the target. 

\begin{table*}[htb]
	\centering
	\caption{Results for c17 and 4-bit S-Box circuits}
	\label{tab:res}
	\begin{threeparttable}

			\begin{tabular}{c c ||c c c c |  c c c c}
				
		\multicolumn{2}{c||}{\multirow{2}{*}{\textbf{conditions}}} & \multicolumn{8}{c}{\textbf{results}}                          \\ \clineB{3-10}{3}
		\multicolumn{2}{c||}{}                  & \multicolumn{4}{c}{\textbf{c17}} & \multicolumn{4}{c}{\textbf{S-Box}} \\  \clineB{1-10}{3}

				fault injection  & probed nodes & CPU time(s)  & iterations & unique? & \begin{tabular}[c]{@{}l@{}}CPU time(s)\\(limited functions)\end{tabular} & CPU time(s)  & iterations & unique? & \begin{tabular}[c]{@{}l@{}}CPU time(s)\\(limited functions)\end{tabular} \ \\ \hhline{--||----|----} 
				\xmark  & \xmark 		    &    timeout\tnote{*}       &   -  & -  &  timeout\tnote{*}  			    & timeout\tnote{*}         &   - & - & timeout\tnote{*} \\
				\xmark  & \cmark 			    &    5.990      &   12  & Yes  & 6.425 & 1029           &   16 & No  &485\\ 
				\cmark  & \xmark 			    &   874.894       &  34   & Yes & 328.623 &timeout\tnote{*}  			    & - & - & timeout\tnote{*}    \\ 
				\cmark  & \cmark                    &   9.178      &  24   & Yes   & 8.086 & 611           &   54 & Yes  &438 \\ \hhline{--||----|----} 
				
			\end{tabular}
		\begin{tablenotes}\footnotesize
			\item[*] Timeout is considered after 16 hours.
		\end{tablenotes}

	\end{threeparttable}
\end{table*}

\section{Conclusions}
\label{sec:conclusions}
In this paper, we have introduced a SAT-based invasive reverse engineering technique that uses probing and fault injection to deobfuscate a circuit. Starting with no knowledge about the gate functions or how they are connected, our approach provides the reverse engineer with a specific set of fault injection and probing experiments to perform on the obfuscated circuit that will allow him to eventually resolve all of its unknown gate functions and connections.
Unlike existing SAT attacks, we show that our approach is able to recover the exact gate-by-gate netlist of the obfuscated circuit.

\textbf{Acknowledgement}: This work has been supported by a grant from the National Science Foundation (NSF) under award CNS-1563829 and by University of Massachusetts, Amherst.



\thispagestyle{empty}


\bibliographystyle{acm}
\bibliography{refs}

\end{document}